\documentclass{iaus}
\usepackage{graphicx}

\title[Galactic PN surveys] 
{The past, present and future of Galactic planetary nebula surveys}

\author[Quentin A. Parker et al.]   
{Quentin A. Parker$^1\,^2\,^3$, D.J. Frew$^1\,^2$, A. Acker$^4$ \& B. Miszalski$^5$}

\affiliation{$^1$Department of Physics and Astronomy, Macquarie University, Sydney, NSW 2109, Australia \\[\affilskip]
$^2$Macquarie University Research Centre in Astronomy, Astrophysics \& Astrophotonics\\[\affilskip]
$^3$Australian Astronomical Observatory, PO Box 296, Epping, NSW
1710, Australia\\email: {\tt quentin.parker@mq.edu.au}\\ [\affilskip]
$^4$ Observatoire Astronomique
de Strasbourg, 11 rue de l' Universit\'e , 67000 Strasbourg, France\\[\affilskip]
$^5$ South African Astronomical Observatory, P.O. Box 9, Observatory, 7935, South Africa}

\pubyear{2012}
\volume{283}  
\pagerange{1--8}
\jname{IAU Symposium 283}
\editors{A. Manchado \& L. Stanghellini, eds.}
\begin{document}

\maketitle

\begin{abstract}
Over the last decade Galactic planetary nebula discoveries have entered a golden age due to the emergence of high sensitivity, high resolution narrow-band surveys of the Galactic plane. These have been coupled with access to complimentary, deep,  multi-wavelength surveys across near-IR, mid-IR and radio regimes in particular from both ground-based and space-based telescopes. These have provided powerful diagnostic and discovery capabilities. In this review these  advances  are put in the context of what has gone before, what we are uncovering now and through the window of opportunity that awaits in the future. The astrophysical potential of this brief but key phase of late stage stellar evolution is finally being realised.
\end{abstract}

\firstsection 
\section{Introduction}

In the five years since the last IAU Symposium PN review (Parker \etal\ 2006a), much progress has been made across a very broad range of planetary nebulae (PN hereafter) research activities. Indeed, there has been so much progress in PN studies and the associated improved insights into late phase stellar evolution, that it is impossible to provide a comprehensive review in the space available of all the major discoveries and their facilitating surveys. Instead we concentrate on our personal impressions of the vigour and importance of key areas of survey progress. These include significant advances in our ability to eliminate PN mimics that have biased previous generations of catalogues, improved distance determinations that have long been one of the most intractable issues in the field,  discovery of  large numbers of new Galactic PN and the uncovering of significant new samples of binary and [WR] central stars.  The future prospects are bright for better understanding of this brief,  complex phase of late-stage stellar evolution across the full range of PN type and evolutionary stage.

\section{The role of PN as astrophysical tools: why we search for them}
PN are among the most fascinating and varied of celestial phenomena originating from  $\sim$1--8~M$\hbox{$\odot$ }$ stars which constitute 90\% of all stars above solar. They dominate the energetics and chemical enrichment process of our entire Galaxy where the PN formation rate should equal the death rate of old, lower mass stars. PN are key probes of nucleosynthesis processes and Galactic abundance gradients and act as powerful tracers of our Galaxy's star-forming history and chemical evolution (Dopita \etal\ 1997; Karakas  \etal\ 2009). PN provide a visible fossil record of the mass loss process off the AGB, including their faint, extensive AGB halos (e.g. Corradi \etal\  2003, Frew \etal\  this meeting) where the physics and mass-loss timescales can be studied. They are useful kinematical probes, visible to large Galactic distances due to their rich emission-line spectra that allow their detection in external galaxies where the bright end fit to their luminosity functions provides a reliable standard candle (e.g. Ciardullo  2010). PN spectra valuably provide expansion velocities, diagnostics of plasma physics, ionization stratification, and, if of high enough quality, abundances. They can also be used to derive the luminosity, temperature and mass of their central stars, provided the important diagnostic emission lines are detected and the central stars (CSPN hereafter) are seen (note only $\sim$25\% of all PN currently have an unequivocally identified CSPN). Furthermore, the integrated flux, surface brightness and radius change in ways that can be predicted by current stellar and hydrodynamic theory. Finally, their observed, complex morphologies provide vital clues to their formation and shaping mechanisms, including the role played by magnetic fields, binary central stars (e.g. Moe \& De Marco 2006, Frank this meeting) or even massive planets (e.g. Soker, 1997, 2006).  These diverse capabilities drive their scientific exploitation and motivate the search for and study of PN in our Galaxy. 

\section{Current PN catalogues \& past problems}
PN are very heterogeneous in their intrinsic and observed properties and may arise via different pathways (see Frew \& Parker this meeting). This causes problems in efforts to compile representative samples for unbiased study. These include their true and projected morphologies  (e.g.  papers by Manchado,  Shaw, this meeting); evolutionary state; angular/physical extent; expansion velocity; variation in nebula excitation and ionisation stratification; central star progenitor mass; chemical composition; surface brightness distribution; presence (and/or detectability) of AGB halos, ansae and external low-ionisation structures. Other issues involve the level of intervening dust and internal/external extinction; degree of interaction  and modification by the surrounding environment as a function both of environment and PN age (e.g. Sabin \etal\ this meeting); the influence of companions (binary or sub-stellar,  e.g. Miszalski \etal\ 2009a, 2009b and this meeting) and magnetic fields  (e.g. Huggins this meeting). There is also the wide variety in the observed properties of PN central stars including the [WR], WELS (e.g. DePew \etal\ 2011) and PG1159 stars. Importantly,  many key characteristics,  including their true physical size and ages, and indeed much of their science capability, depends on knowing distances to reasonable accuracy, while also ensuring that the objects are bona-fide PN!

It is hardly a surprise that previous PN catalogues are incomplete, biased and not representative of the true range of PN types and evolutionary stage. The current status of PN catalogues past and present is summarised in Table~1, updated from  \cite[Parker \etal\ (2006b)]{Parker et al06b} to reflect the further 600 Galactic PN found in the interim. This brings the current total to $\sim3000$. Note this  is still considerably less than the various predictions for the Galactic PN population which vary wildly depending on the estimator employed. For example population synthesis  yields 6600--46,000 PN depending on whether the binary hypothesis for PN formation is required (e.g. De Marco 2009). 
Finally, the local space density estimate derived from the volume limited samples  indicate 13,000 PNe with  r$<$0.9~pc (see table~1 in Jacoby \etal\ 2010). We clearly need to improve the detection completeness for Galactic PN because this number has a direct bearing on stellar evolution, PN formation, chemical enrichment rates, the role of binary stars and the ecology of our Galaxy.  

Apart from the major catalogues there have been some notable supplements by  contributions from other investigations. These include  Boumis \etal\ (2006) who added $\sim100$ PN and the ongoing trawl through over 4000 square degrees of the broad-band Schmidt on-line archives by the Deep Sky Hunters (DSH) consortium (e.g. Jacoby \etal\ 2010 and Kronberger \etal\ this meeting, see Fig.1) where a further 100 valuable, higher Galactic latitude PN have so far been confirmed. More automated techniques to trawl through the existing on-line broad-band imaging archives for PN are also underway such as the ETHOS survey (e.g. Miszalski \etal\, 2011a and this meeting). Nevertheless, taken together this is still a  factor of 2 or 3 short of the lowest estimates for the total Galactic PN population. A key point to remember is the significant numbers of true PN at the very faint, highly evolved end that have been shown to exist in the local volume sample of Frew  \& Parker (2006) but which rapidly become undetectable once distances become greater than a few kpc.
\begin{figure}[bh]
\begin{center}
 \includegraphics[width=4.7in]{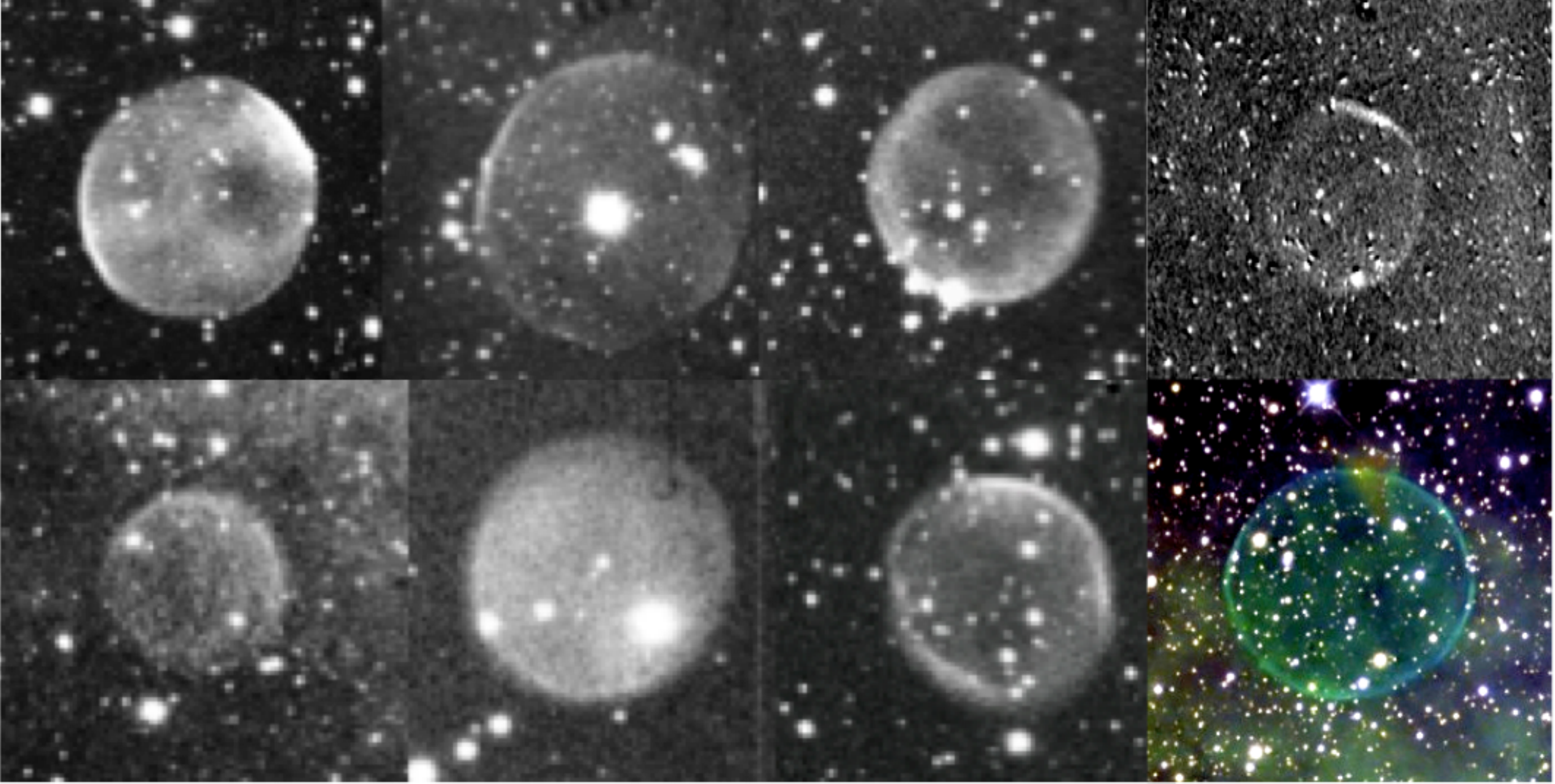} 
 \caption{Six selected DSH PN discoveries (see Kronberger this meeting) and two amateur discoveries by N. Outters  (RHS top) and  D.M. Jurasevitch (RHS bottom). 
A high fraction of round PN have been found in both cases at higher Galactic latitudes surveyed where ISM interactions will be reduced. However,  there may also be other factors at play as bipolar PN are found at lower Galactic scale height too (e.g. Parker \etal\ 2006b).}
   \label{fig1}
\end{center}
\end{figure}

\begin{table}
  \begin{center}
  \caption{Past \& current Major PN Compilations \& Catalogues.}
  \label{tab1}
 {\scriptsize
  \begin{tabular}{lr}\hline 
{\bf Before 2000} &  \\ \hline
Acker \etal\ (1992; 1996)  &  1386 + 489 poss. \\
Kohoutek (2000) updated & 1510 \\ \hline
 {\bf Present major results from new homogeneous surveys} & \\ \hline
MASH I,  Parker \etal\ (2001, 2006b) 	&				$\sim900$ \\
MASH-II, Miszalski \etal\ (2008) 		&		 		$\sim350$\\
Boumis \etal\ 2006 	&				$\sim100$\\
IPHAS, Sabin  \etal\ (2010), Viironen \etal\ (2009ab) \& Corradi \etal\, this meeting	& $\sim500$\\ 
DSH, Jacoby \etal\ (2010), Kronberger \etal\ this meeting			&		$\sim100$\\ \hline
{\bf Current total Galactic PN population known:}	&		$\sim3000$  \\ \hline

  \end{tabular}
  }
 \end{center}
\vspace{1mm}
 \scriptsize{
 {\it Notes}: The Acker \etal\ and Kohoutek compilations cover essentially the same PN and include all historical discoveries from Herschel onwards until the advent of MASH. Also, the Acker \etal\ catalogues addtionally include results of the Acker-Stenholm spectrophotometric survey. All catalogues are now being compiled into a single resource (Miszalski \etal\ this meeting \& Miszalski, Acker, Parker \& Oschenbein,  in preparation).\\}
\end{table}

\section{PN discoveries and characteristics at non-optical wavelengths}
Jacoby \&Van de Steene (2004) undertook an on/off-band bulge survey  at [SIII]~9532\AA, discovering 94 candidate PN.
Also large numbers  of PN candidates have been selected via their IRAS colours but confirmatory success rates have been low (e.g. Suarez \etal\ 2006, Ramos-Larios \etal\ 2009) so this is an inefficient technique.
Mid-IR space-telescope images from Spitzer and WISE now allow detection of very reddened PN which may be invisible optically (eg. Cohen \etal\ 2005, Kwok \etal\ 2008; Phillips \& Ramos-Larios 2008).
Furthermore, Carey \etal\ (2009) and Mizuno \etal\ (2010) have noted 416 compact ($<$1~arcmin) ring, shell and disk-shaped sources in the Galactic plane in 24$\mu$m Spitzer MIPSGAL images. Our belief is that many of these will turn out to be strongly reddened, high-excitation PN with only a minority being circumstellar nebulae around massive stars (Wachter \etal\ 2010). Note that PN can be quite strong mid-IR emitting objects because of PAH emission, fine structure lines, H${_2}$ lines (UWISH2 survey, Froebrich \etal\  2011) and thermal dust emission within the nebulae and in circumnuclear disks. 

\begin{figure}[bh]
\begin{center}
 \includegraphics[width=3.5in]{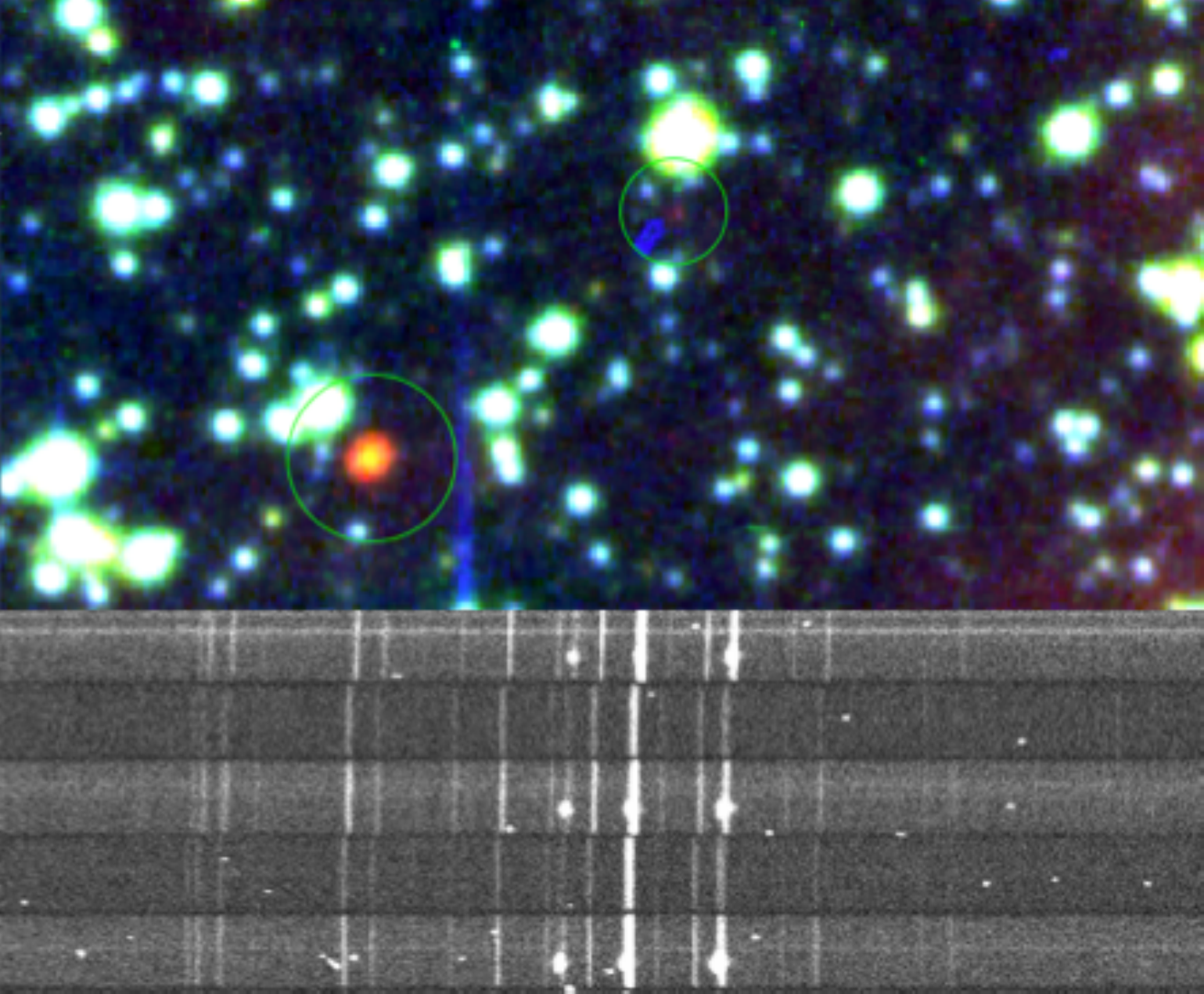} 
 \caption{Upper panel: RGB  colour montage from SuperCOSMOS H$\alpha$, broad-band red and B${_j}$ image of 2$\times$1~arcmin region covering two newly discovered Galactic PN (circled) selected purely on the basis of their GLIMPSE colours. Lower panel: 3$\times$1~arcsecond spectral slices from WiFeS red arm data centred on the brighter newly discovered PN interleaved with sky regions with the nod \& shuffle technique. Wavelength increases from left to right. Note the  [NII]$> $H$\alpha$ ratio from the PN  emission lines  clearly visible as compact knots in three consecutive image slices.}
   \label{fig2}
\end{center}
\end{figure}

Recently, Cohen \etal\ (2007, 2011) analysed optically detected PN and candidates in the GLIMPSE-I survey with the aim to develop robust, multi-wavelength classification and diagnostic tools to provide purer PN samples. The ultimate goal is to recognise PN using only MIR and radio characteristics, enabling the trawl for hidden PN in obscured Galactic regions. In Fig.2 we show the first spectroscopically confirmed PN selected purely on the basis of MIR colours as a proof of concept (Parker \etal\ in prep). Two newly discovered PN are shown (both circled) in the upper panel.  The brighter could have been found by MASH but was missed due to its compactness. The more northerly PN is very faint optically and is highly unlikely to have been picked up visually. We have used the powerful WiFeS image slicer on the ANU 2.3m telescope (Dopita \etal\ 2010) to obtain integrated areal optical spectroscopy to confirm these two PN which, despite  high extinction, have faint traces of [OIII]  in the blue. In the lower panel we show several of the 1~arcsecond slices across the brighter nebula from the WiFeS red arm which clearly show [NII]$>$H$\alpha$ which, when combined with the [OIII] detection, clearly points to a PN origin. Studies of PN in the radio have also recently advanced through the work of Bojicic \etal\ (2011a, 2011b) where targeted ATCA observations of significant numbers of MASH PN were made and where radio surface brightness has been shown to extend naturally to the lower optical limits. There have also been recent advances in the discovery of X-rays from hot bubbles within the central regions of many of the higher surface brightness well known PN (e.g. Guerrero, Chu \& Gruendl 2011, and Guerrero \etal\ this meeting).

\section{The problem of contaminants and their effective elimination}
Other  objects with extended emission  such as HII regions, Str\"omgren zones, supernova remnants, nova shells, bow-shock nebulae, population~I Wolf-Rayet shells and symbiotic systems (e.g. Corradi \etal\  2008) can masquerade as PN (Parker \& Frew, 2010). This has been a major problem undermining the scientific integrity of previous surveys where identification is complicated by the wide variety of morphologies, ionization properties and surface-brightness distributions exhibited by the PN family. Observed PN characteristics reflect stages of nebular evolution, progenitor mass and chemistry and the possible influence of common envelope binaries (Moe \& DeMarco 2006) magnetic fields or even sub-solar planets (Soker 1997, 2006). 

However, we  have recently tested and developed criteria to more effectively eliminate contaminants  only possible because online availability of multi-wavelength surveys and detailed spectroscopic and other data  has enabled clear discrimination tools to be developed. Judicious application of improved multi-wavelength, kinematic and spectroscopic diagnostics (Frew \& Parker 2010) identify imposters from the overall body of evidence and  allow construction of  clean, volume-limited samples to faint luminosities. We can then construct reliable luminosity functions and derive accurate PN space densities and birth rates  to compare with the WD rate. Cohen \etal\ (2011) using these multi-wavelength criteria in the GLIMPSE zone ($|b| \leq$1~deg.)  showed that 45\% of previously known pre MASH PN are HII region contaminants. The MASH contaminant fraction was only 5\% in the same zone because these techniques had already been largely applied to MASH. Note the presence of external filaments, extended structures and/or  amorphous halos in the MIR generally indicates an HII region. Importantly, the observed PN MIR/radio flux ratios and IRAC colour indices appear to be robust attributes, invariant among the many groups of PN. The median PN MIR/radio is 4.7$\pm$1.1 and does not vary significantly with PN evolutionary phase allowing them to be clearly separated  from their major HII region contaminants, regardless of whether they are diffuse or ultra-compact. Furthermore, the observed IRAC colour indices of LMC PN are statistically the same as those of Galactic PN implying that any metallicity differences are not reflected in the MIR PN attributes.

\section{Reliable distances to Galactic PN}
These have been very hard to measure  accurately with various statistical distance scales in use (e.g. Stanghellini, Shaw \& Villaver 2008) but they suffer significant (factor of two or more) errors, both internal and systematic. Crucially, many physical PN characteristics depend on decent distances, including key parameters such as ionised mass, PN size and CSPN luminosity etc.  Primary distance estimators, such as trigonometric parallax, can only be used in the solar vicinity (eg. Benedict  \etal\ 2009).  A  simple, reliable H$\alpha$ surface-brightness radius (SB-r) relation has now been developed that provides more accurate distance estimates for the most evolved PN (Frew \& Parker 2006; 2007). Such a relation could only be derived with the advent of homogeneous H$\alpha$ surveys (eg. Gaustad \etal\  2001; Parker \etal\  2005; Drew \etal\ 2005) and reliable, integrated emission line fluxes (e.g. Kovacevic \etal\ 2011; Bojicic \etal\ this meeting). 
The advantage over previous similar efforts lies in the techniqueÕs large dynamic range, improved ability to effectively remove contaminants and in intelligent use of the very best data. Fuller exploitation of the inherent astrophysical potential of PN is now possible.

\begin{figure}[b]
\begin{center}
 \includegraphics[width=3.6in]{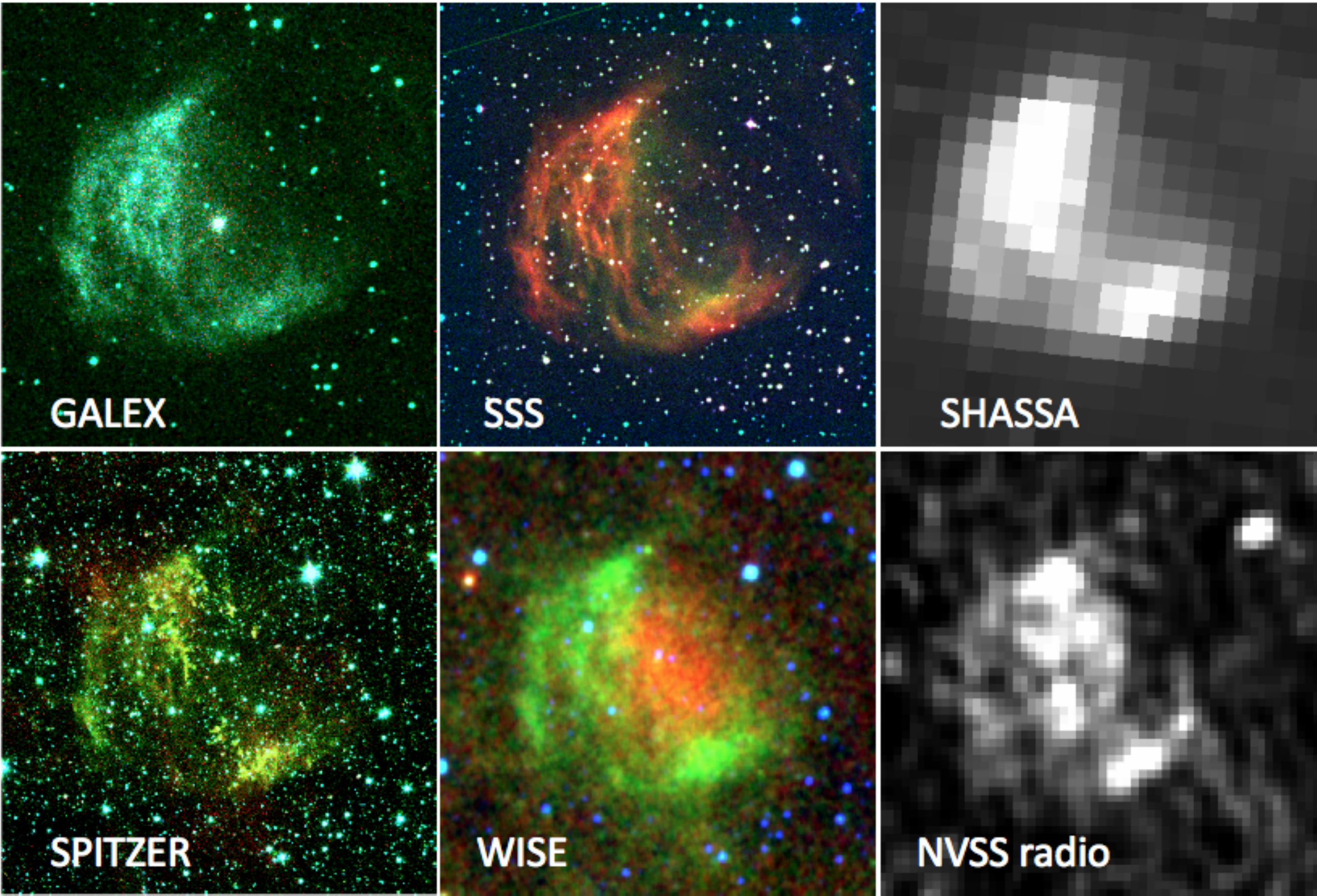} 
 \caption{Multi-wavelength montage of Abell 21. The GALEX  NUV (green) data shows the central ionising star. The optical SSS data is R:red, G:blue \& Blue:I. The green is 5.8$\mu$m in Spitzer. WISE data reveals the inner ionised zone where the MIR [OIV] line, a proxy for HeII in the optical, dominates the 22$\mu$m (red) WISE band where green is 12$\mu$m, blue 4.7$\mu$m)}
   \label{fig3}
\end{center}
\end{figure}

\section{The era of high quality, multi-wavelength imaging surveys: a golden age of PN discovery and catalogue purity}
The advent of deep, high resolution narrow-band optical surveys revolutionised our ability to trawl for new PN. In particular
the recent H$\alpha$ surveys of the southern and northern Milky Way by Parker \etal\  (2005) and Drew \etal\ (2005) led to a doubling of known Galactic PN over the last 10 years.  This has mainly been through the Macquarie/AAO/Strasbourg H$\alpha$ MASH catalogues of Parker \etal\ (2006b) and Miszalski \etal\ (2008) which listed $\sim$1250 true, likely and possible new southern Galactic plane PN. These are now being supplemented by the on-going discoveries from the IPHAS survey in the northern plane reported by Viironen \etal\ (2009a,b) and Sabin \etal\ (2010 and this meeting), where more than 500 new PN candidates have been uncovered with over 100 so far spectroscopically confirmed (Sabin \etal\, in preparation).  These discovery data can be  coupled with the major ground and space-based radio (NVSS, MOST), multi-wavelength optical (e.g. SDSS), NIR (2MASS, UKIDSS, VVV) and MIR surveys (MSX, GLIMPSE, MIPSGAL, WISE, Herschel) of our Galaxy.  Where available, GALEX UV data can also be used to reveal hitherto invisible PN ionising stars (e.g. Frew \etal\ 2011 and see Fig.1) while more extensive high resolution X-ray data is also being obtained with Chandra. Taken together these surveys have further enhanced the PN discovery potential in new ways (e.g. Mizalski  \etal\  2011b) and have provided fresh insights into their multi-wavelength characteristics (e.g. Cohen \etal\  2007, 2011) while permitting the more robust elimination of contaminants that have significantly impacted the integrity of previous catalogues and the scientific analysis of their content (Frew \& Parker 2010;  Boissay \etal\ this meeting).  As an example, a multi-wavelength montage of  Abell 21 is shown in Fig.3. Note the GALEX UV data reveals the ionising star whereas the WISE data reveals the inner ionised zone where the MIR [OIV] line (red in this 22~$\mu$m IRAS band) is a proxy for HeII in the optical in this case, and not heated dust.

\section{The Future}
The MASH and IPHAS surveys of Galactic PN are  proving to be a  fruitful basis for much on-going research into the properties of the global PN population. This includes the provision of  reliable, homogeneous H$\alpha$, [OIII] and other line fluxes for large numbers of Galactic PN for the first time (e.g. Kovacevic \etal\, 2011; Frew, Bojicic \& Parker, in prep and Bojicic \etal\ this meeting). 
Programs to investigate quantitative differences in  multi-wavelength characteristics and how these relate to PN age, type and chemistry are underway as are attempts to derive mass-loss estimates and other key PN physical parameters via searches for PN haloes, ansae and low-ionisation structures. We are also searching for PN in open clusters (e.g. Parker \etal\ 2011, Frew \etal\ in prep) and continuing our searches for emission-line CSPN including those with [WR] and WELS, PG1159 stars (e.g. DePew \etal\ 2011). Powerful new observational techniques and advances in technology (e.g. IFUs and image slicers) provide 3-D data cubes which permit more efficient, detailed, spatio-kinematic study of PN compared with traditional long-slit techniques, PV (position-velocity) diagrams and modeling via `SHAPE' analysis (e.g. Steffen \& L\'opez,  2006). This is especially true of low surface brightness PN which are otherwise very difficult to study properly. Apart from the continued development of IFUs to larger formats of particular relevance for PN studies are fibre hexabundles (e.g. Bland-Hawthorn \etal\  2010) and OH suppression fibres for the NIR (e.g. Ellis \& Bland-Hawthorn 2008).

We are now in the multi-wavelength, large-scale survey era exemplified by the ESO public surveys on the VISTA and VST telescopes such as VPHAS+, VVV  and future surveys with Skymapper, LSST and the Dark energy Camera on the Blanco 4~m. The powerful discovery and evaluation techniques developed will provide a more complete Galactic PN  inventory, revealing  the true underlying PN population. Questions we hope to answer before the next symposium include resolving the role of binarity, e.g. by a CSPN study of a highly complete local volume sample, uncovering the drivers of morphology and establishing the role of dust, especially in terms of ejected mass budget. A fuller description of the evolutionary pathways that lead to PN  (Frew \& Parker this meeting), an understanding of the PNLF, especially at the key bright end, from highly complete Bulge and local volume samples should also be possible. More complete identifications and  studies of the heterogeneous character of central stars are also needed (e.g. Weidmann \& Gamen, 2011), including the role of [WR]  CSPN, the possible emergence of a dynamical age sequence, scale-height segregation and the recent discovery of [WN] and [WNC] CSPN which are not meant to exist (e.g. Todt \etal\ 2010, DePew \etal\  2011). Finally, there is an urgent need for more detailed abundances as only  $\sim$150 accurate Galactic PN abundances are currently available from the known population of $\sim3000$ (5\%).


\begin{thebibliography}{}

\bibitem [Acker \etal\ (1992)]{Acker etal92}
{Acker, A., \etal\ } 1992, 1996,
\textit{Strasbourg-ESO Catalogue of Galactic PNe \& 1st suppl.}



\bibitem[Benedict, G., \etal\ (2009)]{Benedict etal09}
{Benedict, G.,  \etal\ } 2009,
\textit{AJ}, 138, 1969
 
\bibitem[Bland-Hawthorn,J., \etal\ (2010) ]{Bland10}
{Bland-Hawthorn, J., \etal\ } 2010, 
\textit{Opt. Exp.}, 19, 2649


\bibitem[Bojicic., I., \etal\ (2011)]{Bojicic11a}
{Bojicic, I., Parker, Q.A., Filipovic, M.D., \& Frew, D.J., } 2011a,
\textit{MNRAS}, 412, 223

\bibitem[Bojicic., I., \etal\ (2011)]{Bojicic11b}
{Bojicic, I., Parker, Q.A., Frew, D.J., \etal\ } 2011b,
\textit{Astron. Nachr}, 332, 697

\bibitem[Boumis \etal\ (2003)]{Boumis etal03} 
{Boumis, P., \etal\ } 2003,
\textit{MNRAS}, 339, 735

\bibitem[Boumis \etal\ (2006)]{Boumis etal06} 
{Boumis, P., \etal\ } 2006,
\textit{MNRAS}, 367, 1551

\bibitem[Carey, S.J., \etal\ (2009)]{Carey etal09}
{Carey, S.J., \etal\ } 2009,
\textit{PASP}, 121, 76

\bibitem[Ciardullo, R., (2010)]{Ciardullo etal10}
{Ciardullo, R., } 2010,
\textit{PASA}, 27, 149

\bibitem[Cohen, M.C., \etal\  (2005)]{Cohen etal05}
{Cohen, M.C., \etal\ } 2005, 
\textit{ApJ}, 627, 446

\bibitem[Cohen, M.C.,  \etal\ (2007)]{Cohen etal07}
{Cohen, M.C., Parker, Q.A., \etal\ } 2007, 
\textit{MNRAS}, 374, 979

\bibitem[Cohen, M.C.,  \etal\  (2011)]{Cohen etal11}
{Cohen, M.C., \etal\ } 2011,
\textit{MNRAS}, 413, 514

\bibitem[Corradi \etal\ (2003)]{Corradi etal03}
{Corradi, R.,  \etal\ } 2003,
\textit{MNRAS}, 340, 417

\bibitem[Corradi \etal\ (2008)]{Corradi etal08}
{Corradi, R.,  \etal\ } 2008,
\textit{A\&A}, 480, 409

\bibitem[De Marco., O (2009)]{DeMarco 09}
{De Marco, O.,} 2009,
\textit{PASP}, 121, 316

\bibitem[DePew, K.,  \etal\ (2011)]{DePew etal11}
{DePew, K., Parker, Q.A., Miszalski, M., \etal\ } 2011,
 \textit{MNRAS}, 413, 2812
 
 
\bibitem[Dopita,M.,  \etal\ (1997)]{Dopital etal97}
{Dopita, M., \etal\ } 1997,
\textit{ApJ}, 474, 188
 
\bibitem[Dopita, M., \etal\  (2010)]{Dopita etal10}
{Dopita, M., \etal\ } 2010,
\textit{Ap\&SS}, 327, 245

\bibitem[Drew, J., \etal\ (2005)]{Drew etal05}
{Drew, J., \etal\ } 2005,
 \textit{MNRAS}, 362, 753

\bibitem[Ellis, S. \& Bland-Hawthorn, J., (2008)]{Ellis08}
{Ellis, S. \& Bland-Hawthorn, J.,} 2008,
\textit{MNRAS}, 386, 47


\bibitem[Frew, D.J., \& Parker, Q.A., (2006)]{Frew06}
{Frew, D.J, Parker, Q.A.,} 2006, 
\textit{IAU Symp. 234}, 49

\bibitem[Frew, D.J., \& Parker, Q., (2007)]{Frew07}
{Frew, D.J., Parker, Q.A.,} 2007, 
\textit{APN-IV}, p.475

\bibitem[Frew, D.J., \& Parker, Q.A., (2010)]{Frew10}
{Frew, D.J., \& Parker, Q.A.,} 2010, 
\textit{PASA}, 27, 129

\bibitem[Frew, D.J., \etal\  (2011)]{Frew etal11}
{Frew, D.J., \etal\ } 2011, 
\textit{PASA}, 28, 83

\bibitem[Froebrich, D., \etal\ (2011)]{Froebrich etal11}
{Froebrich, D., \etal\  } 2011.
\textit{MNRAS}, 413, 480

\bibitem[Gaustad, J., \etal\ (2001) ]{Gaustad01}
{Gaustad, J., \etal\ } 2001, 
\textit{PASP}, 113, 1326

\bibitem[Guerrero, M., Chu, You-Hua., Greundl, R.A., (2011)]{Gerrero etal11}
{Guerrero, M., Chu, You-Hua., Greundl, R.A.,} 2011,
\textit{The X-ray Universe, Berlin}, 198


\bibitem[Jacoby, G., \&Van de Steene, G.,  (2004)]{Jacoby04}
{Jacoby, G., \& Van de Steene, G.,} 2004,
\textit{A\&A}, 419, 563 
 
\bibitem[Jacoby, G., \etal\ (2010)]{Jacoby etal10}
{Jacoby, G., \etal\ } 2010,
\textit{PASA}, 27, 156


\bibitem[Karakas, A., \etal\ (2009]{Karakas etal09}
{Karakas, A., \etal\ } 2009,
\textit{ApJ}, 690, 1130


\bibitem[Kovacevic, A., Parker, Q.A., Jacoby, G., \etal\  (2011)]{Kovacevic etal11}
{Kovacevic, A., Parker, Q.A., Jacoby., G  \etal\ } 2011,
\textit{MNRAS}, 414, 860 


\bibitem[Kohoutek, L., (2001)]{Kohutek01}
{Kohoutek, L., } 2001,
\textit{Catalogue of Galactic PN (updated version 2000)}

\bibitem[Kwok, S., \etal\   (2008)]{Kwok etal08}
{Kwok, S., Zhang, Y., Koning, N., Huang, H.-H., Churchwell, E.,} 2008,
\textit{ApJS}, 174, 426


\bibitem[Miszalski, B.,  \etal\  (2008)]{Miszalski etal08}
{Miszalski, B., \etal\ } 2008,
\textit{MNRAS}, 384, 525 

\bibitem[Miszalski, B.,  \etal\  (2009a)]{Miszalski etal09}
{Miszalski, B., \etal\ } 2009a,
\textit{A\&A}, 496, 813 

\bibitem[Miszalski, B.,  \etal\  (2009b)]{Miszalski etal09}
{Miszalski, B., \etal\ } 2009b,
\textit{A\&A}, 505, 249 


\bibitem[Miszalski, B.,   \etal\  (2011)]{Miszalski etal11a}
{Miszalski, B.,  \etal\ } 2011a,
\textit{MNRAS}, 413, 1264 


\bibitem[Miszalski, B.,   \etal\  (2011)]{Miszalski etal11b}
{Miszalski, B.,  \etal\ } 2011b,
\textit{A\&A}, 531, 157 

\bibitem[Mizuno \etal\  (2010)]{Mizuno etal10}
{Mizuno, D.R., \etal\ } 2010,
\textit{AJ}, 139, 1552

\bibitem[Moe,M., \& De Marco, O., (2006)]{Moe06}
{Moe, M \& De Marco, O.,} 2006,
\textit{ApJ}, 650, 916

\bibitem[Parker \etal\ (2005)]{Parker_etal05}
{Parker, Q.A., Phillipps, S., Pierce, M.J., \etal\, } 2005,
\textit{MNRAS}, 362, 689 

\bibitem[Parker \etal\ (2006)]{Parker etal06a}
{Parker, Q.A., Acker, A., Frew, D.J., \& Reid, W.A.,} 2006a,
\textit{IAU Symp. 234}, 1

\bibitem[Parker \etal\ (2006)]{Parker_etal06b}
{Parker, Q.A., Acker, A., Frew, D.J., \etal\ } 2006b,
\textit{MNRAS}, 373, 79 

\bibitem[Parker \etal\ (2011)]{Parker_etal11}
{Parker, Q.A., Frew, D.J., Miszalski, B., \etal\ } 2011,
\textit{MNRAS}, 413, 1835 

\bibitem[Phillips, J.P., Ramos-Larios, G. (2008)]{Phillips08} 
{Phillips, J.P. \& Ramos-Larios, G.,} 2008, 
\textit{MNRAS}, 386, 995

\bibitem[Ramos-Larios, G., \etal\ (2009)]{Ramos09}
{Ramos-Larios, G., \etal\ } 2009, 
\textit{A\&A}, 501, 1207

\bibitem[Sabin, L., \etal\ (2010)]{Sabin10}
{Sabin, L., \etal\ } 2010,
\textit{PASA}, 27, 166

\bibitem[Soker, N (1997)]{Soker97}
{Soker, N.,} 1997, 
\textit{ApJS}, 112, 487

\bibitem[Soker., N (2006)]{Soker06}
{Soker, N.,} 2006, 
\textit{ApJ}, 640, 966

\bibitem[Stanghellini, L., Shaw, R.A. \& Villaver, E., (2008)]{Stanghellini etal08}
{Stanghellini, L., Shaw, R.A. \& Villaver, E., } 2008,
\textit{ApJ}, 689, 194

\bibitem[Steffen, W. \& L\'opez, J.A., (2006)]{Stefen06}
{Steffen, W. \& L\'opez, J.A.,} 2006,
\textit{RMxAA}, 42, 99

\bibitem[Suarez \etal\ (2006)]{Suarez etal06}
{Suarez, O., \etal\, }2006,
\textit{A\&A}, 458, 173

\bibitem[Todt \etal\ (2010)]{Todt etal10}
{Todt, H., \etal\, } 2010,
\textit{A\&A}, 515, 83

\bibitem[Viironen \etal\ (2009a)]{Viironen etal09}
{Viironen, K., \etal\ } 2009a, 
\textit{A\&A}, 502, 113

\bibitem[Viironen \etal\ (2009b)]{Viironen etal09}
{Viironen, K., \etal\ } 2009b, 
\textit{A\&A}, 504, 291

\bibitem[Weidmann \& Gamen (2011)]{Wiedmann11}
{Wiedmann, W.A., \& Gamen, R.} 2011,
\textit{A\&A}, 526, A6

\bibitem[Wachter \etal\  (2010)]{Wachter etal10}
{Wachter, S., \etal\ } 2010,
\textit{AJ}, 139, 2330

\end{thebibliography}
\end{document}